# Generalized Penner models to all genera

*J. Ambjørn*

The Niels Bohr Institute

Blegdamsvej 17, DK-2100 Copenhagen Ø, Denmark

*C. F. Kristjansen*

NORDITA

Blegdamsvej 17, DK-2100 Copenhagen Ø, Denmark

*Yu. Makeenko*

The Niels Bohr Institute

Blegdamsvej 17, DK-2100 Copenhagen Ø, Denmark

## Abstract

We give a complete description of the genus expansion of the one-cut solution to the generalized Penner model. The solution is presented in a form which allows us in a very straightforward manner to localize critical points and to investigate the scaling behaviour of the model in the vicinity of these points. We carry out an analysis of the critical behaviour to all genera addressing all types of multi-critical points. In certain regions of the coupling constant space the model must be defined via analytical continuation. We show in detail how this works for the Penner model. Using analytical continuation it is possible to reach the fermionic 1-matrix model. We show that the critical points of the fermionic 1-matrix model can be indexed by an integer, $m$, as it was the case for the ordinary hermitian 1-matrix model. Furthermore the $m$'th multi-critical fermionic model has to all genera the same value of $\gamma_{str}$ as the $m$'th multi-critical hermitian model. However, the coefficients of the topological expansion need not be the same in the two cases. We show explicitly how it is possible with a fermionic matrix model to reach a $m = 2$ multi-critical point for which the topological expansion has alternating signs, but otherwise coincides with the usual Painlevé expansion.

# 1   Introduction

The hermitian 1-matrix model with a polynomial potential is by now completely understood. All critical points have been localized and classified. The different types of critical behaviour are indexed by an integer, $m$, and the $m$'th multi-critical model is characterized by $\gamma_{str}$ taking the value $-\frac{1}{m}$ [1]. Furthermore it is well known that if one considers a potential consisting of a linear plus a logarithmic term one can reach a critical point for which $\gamma_{str} = 0$ and for which logarithmic scaling violations occur at genus zero [2]. This is the critical point of the Penner model [3]. The possibility of multi-critical behaviour for generalized Penner models, i.e. models with a logarithmic term and a higher degree polynomial was pointed out in references [4, 5, 6]. However, only the genus zero and the genus one behaviour was addressed.

In reference [7] a complete description of the genus expanded 1-cut solution to the hermitian 1-matrix model with a generic polynomial potential was given. Here we generalize this description to the case where in addition a non polynomial term, namely a logarithm, appears in the interaction. The solution is presented in a form which allows us in a straightforward manner to localize all critical points and to investigate the scaling behaviour of the model in the vicinity of these points. All types of multi-critical points can be addressed and the analysis can be carried out for any genus. The outcome of our analyses is that no other values of $\gamma_{str}$ than $\gamma_{str} = 0$ or $\gamma_{str} = -\frac{1}{m}$, $m = 2, 3, \ldots$ are possible.

In some regions of its coupling constant space the generalized Penner model can only be defined via analytical continuation. We show in detail how this procedure works for the Penner model itself extending the analyses of references [4, 5]. By analytical continuation it is possible to reach the fermionic 1-matrix model [8]. As an application of our study of the generalized Penner model we show that the possible types of multi-critical behaviour for the fermionic 1-matrix model can, as in the hermitian case, be indexed by an integer, $m$. Futhermore the $m$'th multi-critical fermionic model has to all genera the same value of $\gamma_{str}$ as the $m$'th multi-critical hermitian matrix model. However, the coefficients of the terms in the topological expansion will not necessarily be the same for a $m$'th multi-critical point obtained from a hermitian and a fermionic matrix model respectively. For example it is possible in the fermionic case to find a $m = 2$ multi-critical point for which the topological series has alternating signs but otherwise coincides with the usual Painlevé expansion. Having alternating signs, this series might very well be Borel summable.

## 2 The model and its loop equations

The generalized Penner model is defined by the partition function

$$Z = e^{N^2 F} = \int_{N \times N} d\phi \exp(-N \operatorname{Tr} [U(\phi)]) \tag{2.1}$$

where the integration is over hermitian $N \times N$ matrices and

$$U(\phi) = V(\phi) + t \log \phi, \qquad V(\phi) = \sum_{j=1}^{\infty} \frac{g_j}{j} \phi^j. \tag{2.2}$$

We define expectation values in the usual way and introduce the generating functional for 1-loop averages by

$$W(p) = \langle \frac{1}{N} \operatorname{Tr} \frac{1}{p - \phi} \rangle. \tag{2.3}$$

With the normalization chosen above the genus expansion of the free energy and the 1-loop correlator read

$$F = \sum_{g=0}^{\infty} \frac{1}{N^{2g}} F_g, \qquad W(p) = \sum_{g=0}^{\infty} \frac{1}{N^{2g}} W_g(p). \tag{2.4}$$

The loop equations of the model express the invariance of its partition function under field redefinitions. To derive the loop equations in an appropriate form it is convenient to consider the following transformation of the field, $\phi$

$$\phi \longrightarrow \phi + \epsilon \sum_{n \geq 1} \frac{\phi^n}{p^{n+1}} = \phi + \epsilon \left( \frac{\phi}{p(p-\phi)} \right). \tag{2.5}$$

The reason why we do not include a $n = 0$ term in the sum above is that we wish to avoid the appearance of terms of the type $\operatorname{Tr} \phi^{-1}$ from the variation of $\log \phi$ in the action. Introducing the shift of $\phi$, (2.5) in (2.1) we get to the first order in $\epsilon$

$$\int d\phi \left\{ \left( \operatorname{Tr} \left( \frac{1}{p-\phi} \right) \right)^2 - N \operatorname{Tr} \left( U'(\phi) \frac{\phi}{p(p-\phi)} \right) \right\} e^{-N \operatorname{Tr}(U(\phi))} = 0. \tag{2.6}$$

We now introduce as usual corresponding to the matrix $\phi$ an eigenvalue density $\rho(\lambda)$. With our interaction it is necessary to require that the support of the eigenvalue density does not cross the branch cut of $\log \lambda$. Then the equation (2.6) can be written as

$$\frac{1}{p} \oint_C \frac{d\omega}{2\pi i} \frac{\omega V'(\omega)}{p - \omega} W(\omega) + \frac{t}{p} W(p) = (W(p))^2 + \frac{1}{N^2} \frac{d}{dV(p)} W(p) \tag{2.7}$$

where $\frac{d}{dV(p)}$ is the loop insertion operator

$$\frac{d}{dV(p)} \equiv -\sum_{j=1}^{\infty} \frac{j}{p^{j+1}} \frac{d}{dg_j} \tag{2.8}$$

and where $C$ is a curve which encloses the support of the eigenvalue density but which does not cross the branch cut of the logarithm, neither encloses the point $\omega = p$. It is easy to see that (2.7) can be written in the form

$$\oint_C \frac{d\omega}{2\pi i} \frac{\{V'(\omega) + \frac{t}{\omega}\}}{p - \omega} W(\omega) = (W(p))^2 + \frac{1}{N^2} \frac{d}{dV(p)} W(p) \tag{2.9}$$

provided $W(p)$ fulfills the following equation

$$\oint_C \frac{d\omega}{2\pi i} V'(\omega) W(\omega) - t W(0) = 0. \tag{2.10}$$

For $t = 0$ the validity of the relation (2.10) is ensured by the invariance of the partition function under field redefinitions of the type $\phi \to \phi + \epsilon$. For $t \neq 0$ we can also formally derive the relation (2.10) with $W(0) = \langle \frac{1}{N} \text{Tr } \phi^{-1} \rangle$ by considering the same transformation of the field. However, the ill defined quantity $\text{Tr } \phi^{-1}$ appears. By equating the $\frac{1}{p}$ terms in (2.9) we see that a $W(p)$ which fulfills (2.9) will automatically fulfill the requirement (2.10). Hence a solution of (2.9) is automatically a solution of (2.7). The loop equation (2.9) is of exactly the same form as the one of the hermitian 1-matrix model without a logarithmic interaction term. In reference [7] an iterative procedure for solving the latter loop equation was developed and explicit results for $W_g(p)$ as well as $F_g$ for $g = 1$ and $g = 2$ were given. Furthermore the general structure of $F_g$ and $W_g(p)$ for any $g$ was described. From the results of reference [7] we can read off the complete perturbative one-cut solution of the model (2.1).

## 3 The genus zero solution

### 3.1 General results

With the assumption that the singularities of $W(p)$ consist only of one square root branch cut (which in the case of $W(p)$ being a solution of (2.9) is equivalent to $\rho(\lambda)$ having support only on one arc in the complex plane [9]) and with the normalization $W(p) \to 1/p$ as $p \to \infty$, the genus zero contribution to $W(p)$ for the model (2.1) can be written as

$$W_0(p) = \frac{1}{2} \oint_C \frac{d\omega}{2\pi i} \frac{\{V'(\omega) + \frac{t}{\omega}\}}{p - \omega} \left\{ \frac{(p-x)(p-y)}{(\omega-x)(\omega-y)} \right\}^{1/2} \tag{3.1}$$

where $x$ and $y$ are determined by the boundary conditions

$$B_1(x,y) \equiv \oint_C \frac{d\omega}{2\pi i} \frac{\{V'(\omega) + \frac{t}{\omega}\}}{\sqrt{(\omega-x)(\omega-y)}} = 0, \tag{3.2}$$

$$B_2(x,y) \equiv \oint_C \frac{d\omega}{2\pi i} \frac{\{\omega V'(\omega) + t\}}{\sqrt{(\omega-x)(\omega-y)}} = 2. \tag{3.3}$$

It is easy to verify by direct calculation using (3.1) that the condition (2.10) is indeed satisfied for the genus zero solution due to the boundary conditions (3.2) and (3.3). With $V(\phi)$ given by (2.2) the boundary equations can be written as a set of algebraic equations. For $\deg(V(\phi)) = D$ the degree of the boundary equations is $D$ if $t = 0$ and $2D$ if $t \neq 0$. Hence for $t = 0$ an explicit solution can be found for $\deg(V(\phi)) \leq 4$ whereas for $t \neq 0$ an explicit solution can be found only in the case of a linear or a quadratic potential. It is easy to see that for $t = 1$, $x = y = z$ is a possible solution of the boundary equations for any potential. Namely setting $x = y = z$ in the boundary equations (3.2) and (3.3) we find

$$B_1(z,z) = V'(z) - \frac{t}{z}, \qquad B_2(z,z) = z\,V'(z) + t. \tag{3.4}$$

Hence with $t = 1$ and $B_1(z,z) = 0$ the boundary equation $B_2(z,z) = 2$ is automatically fulfilled. Conversely it follows that the only way in which we can have $x = y = z$ is by $t$ being equal to 1. As noted in reference [4] for $t = 1$ we have $W_0(p) = \frac{1}{p}$ independently of $V(\phi)$. This result follows easily from formula (3.1). Another case for which some information can be extracted from the boundary equations even for a generic potential is the case $t = 2$. For $t = 2$ one finds that if the potential $V(\phi)$ is of a definite parity $x$ and $y$ will lie on the imaginary axis. If $V(\phi)$ is even $x$ and $y$ will both be situated either on the negative or on the positive imaginary axis. If $V(\phi)$ is odd one has $x = -y$. By solving the boundary equations we determine the branch points of $W(p)$. However, this does not fix the position of the cut of $W(p)$, i.e. the support of the eigenvalue distribution. The missing information is encoded in the function $G(\lambda)$ [10]

$$G(\lambda) = \int_y^\lambda \left(U'(p) - 2W_0(p)\right) dp. \tag{3.5}$$

The support of the eigenvalue distribution is an arc connecting $y$ to $x$ along which $G(\lambda)$ is purely imaginary and which is embedded in a region where $\mathrm{Re}(G) < 0$. We are used to considering a model of the type (2.1) as meaningful if the branch points $x$ and $y$ are real and the cut lies along the real axis. There are regions of the coupling constant space where such a situation can not be realized. In some of these regions it is still possible to attribute a meaning to the model by analytical continuation. The eigenvalue density is defined according to the rules given above and is now a curve in the complex plane. We note that one might obtain by this prescription a complex valued partition function. A lot of information about the model (2.1) can be extracted without knowing the precise location of the branch points or the cut of $W(p)$. For instance all critical points can be easily localized and classified. All we have to assume is that the branch cut of $W(p)$ does not cross the branch cut of the logarithm. When working with contour integrals this assumption turns into the assumption that the contour does not cross the branch cut of the logarithm.

By means of the solution (3.1) we can calculate the genus zero contribution to the susceptibility, $\chi_0$

$$\chi = -\frac{d^2}{dt^2}F. \tag{3.6}$$

By differentiating (2.1) one gets

$$\chi = \frac{d}{dt}\langle\frac{1}{N}\text{Tr log }\phi\rangle = \oint_C \frac{d\omega}{2\pi i}\log\omega\,\frac{dW(\omega)}{dt}. \tag{3.7}$$

From the expression (3.1) one obtains by direct differentiation using Eqs. (3.2) and (3.3)

$$\begin{aligned}\frac{dW_0(p)}{dt} &= \oint_C \frac{d\omega}{4\pi i}\frac{1}{(p-\omega)\omega}\left(\frac{(\omega-x)(\omega-y)}{(p-x)(p-y)}\right)^{1/2} \\ &= \frac{1}{2}\left\{\frac{1}{p}-\frac{1}{p}\left(\frac{xy}{(p-x)(p-y)}\right)^{1/2}-\frac{1}{((p-x)(p-y))^{1/2}}\right\}.\end{aligned} \tag{3.8}$$

Substituting into Eq. (3.7) and compressing the contour $C$ to the cut of the logarithm, we get

$$\begin{aligned}\chi_0 &= -\frac{1}{2}\int_0^\infty dp\left\{\frac{1}{p}-\frac{1}{p}\left(\frac{xy}{(p-x)(p-y)}\right)^{1/2}-\frac{1}{((p-x)(p-y))^{1/2}}\right\} \\ &= -\frac{1}{2}\log\left\{\frac{\left(2(xy)^{1/2}-(x+y)\right)^2}{16xy}\right\}.\end{aligned} \tag{3.9}$$

We stress that this formula holds for a generic potential.

## 3.2 The Penner model

Here we show how the procedure of analytical continuation works in a simple case, namely the case of the Penner model

$$g_1 = -1, \quad g_i = 0, \ i > 1. \tag{3.10}$$

For this model the boundary equations read

$$B_1(x,y) = -1 - \frac{t}{(xy)^{1/2}} = 0, \tag{3.11}$$

$$B_2(x,y) = -\frac{1}{2}(x+y) + t = 2 \tag{3.12}$$

and we note that we recover from (3.9) immediately the well known result [4]

$$\chi_0 = -\log(1-t) - \log t \tag{3.13}$$

Solving (3.11) and (3.12) for $x$ and $y$ gives

$$x = (t-2) + 2(1-t)^{1/2}, \tag{3.14}$$
$$y = (t-2) - 2(1-t)^{1/2}. \tag{3.15}$$

The analyticity structure of $W(p)$ and hence the location of the support of the eigenvalue distribution is quite different for different values of $t$. A detailed analysis of the analyticity structure based on the function $G(\lambda)$, which can be calculated explicitly, is carried out in Appendix A. Here we just summarize the results.

**t < 0:** The branch points $x$ and $y$ are real and negative and the cut of $W(p)$ is the straight line connecting $x$ and $y$. This position of the cut of $W(p)$ implies that $(xy)^{1/2} = |xy|^{1/2}$ which is in accordance with equation (3.11). The branch cut of the logarithm can for instance be placed along the negative real axis (figure 1). In the limit $t \to 0_-$ the cut of $W(p)$ collides with the cut of the logarithm (figure 2).

**0 < t < 1:** The branch points $x$ and $y$ are again situated on the negative real axis. However, the cut of $W(p)$ now looks as in figure 3. We note that the presence of the loop implies that $(xy)^{1/2} = -|xy|^{1/2}$ which is again in accordance with equation (3.11). For $t \to 1_-$ the branch points $x$ and $y$ collide (figure 4).

**t > 1:** The branch points $x$ and $y$ get an imaginary part and are each others complex conjugate. The cut of $W(p)$ is an arc which connects $y$ and $x$ and which intersects the real axis on its positive part (figure 5). In particular for $t = 2$ the model (2.1) can be interpreted as a fermionic matrix model [8]. We will return to this point in section 6.

## 4 The solution to higher genera

Explicit solutions for $W_1(p)$ and $W_2(p)$ as well as the general structure of $W_g(p)$ for $g > 1$ can be read off from reference [7] and it is easy to show that $W(p)$ fulfills the requirement (2.10) to all genera. From the analysis of reference [7] we also have a detailed knowledge of the free energy of the model. To express the higher genera contributions to $F$ (as well as to any multi-loop correlator) it is convenient to introduce instead of the coupling constants $(\{g_i\}, t)$ a set of moments $\{M_k, J_k\}$

$$M_k = \oint_C \frac{d\omega}{2\pi i} \frac{\{V'(\omega) + \frac{t}{\omega}\}}{(\omega - x)^{k+1/2} (\omega - y)^{1/2}}, \qquad k \geq 1, \qquad (4.1)$$

$$J_k = \oint_C \frac{d\omega}{2\pi i} \frac{\{V'(\omega) + \frac{t}{\omega}\}}{(\omega - x)^{1/2} (\omega - y)^{k+1/2}} \qquad k \geq 1. \qquad (4.2)$$

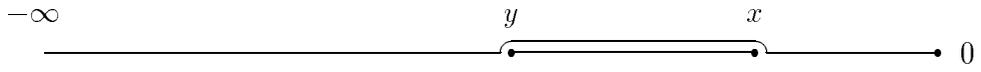

Figure 1: The support of the spectral density for $t < 0$ (the bold line) and the branch cut of the logarithm (the thin line).

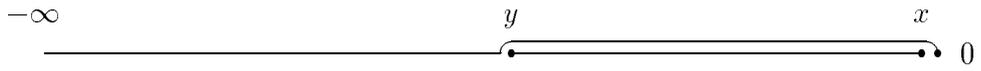

Figure 2: The support of the spectral density for $t \to 0_-$ (the bold line) and the branch cut of the logarithm (the thin line).

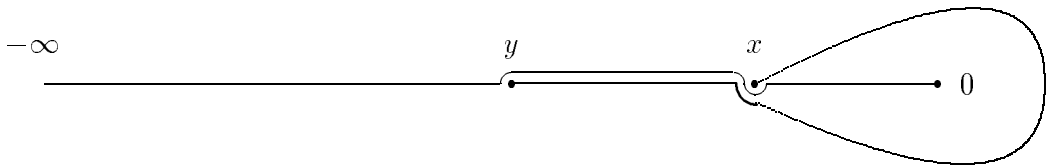

Figure 3: The support of the spectral density for $0 < t < 1$ (the bold line) and the branch cut of the logarithm (the thin line).

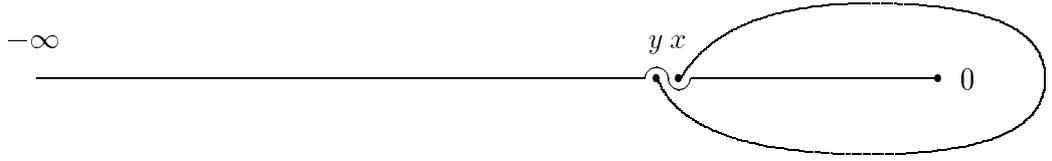

Figure 4: The support of the spectral density for $t \to 1_-$ where $x \to y$ (the bold line) and the branch cut of the logarithm (the thin line).

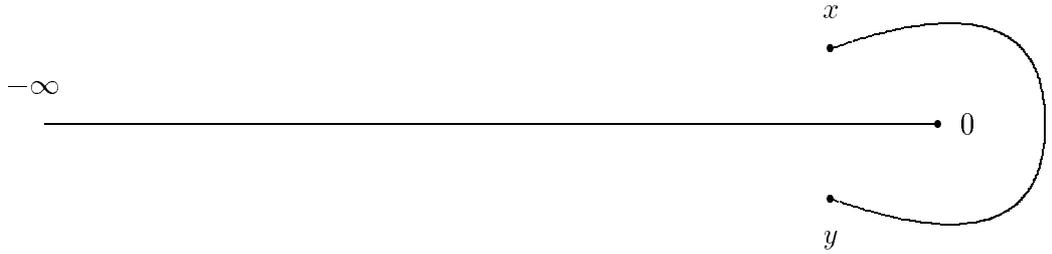

Figure 5: The support of the spectral density for $t > 1$ (the bold line) and the branch cut of the logarithm (the thin line).

The genus one contribution to $F$ reads

$$F_1 = -\frac{1}{24} \ln M_1 - \frac{1}{24} \ln J_1 - \frac{1}{6} \ln d. \qquad (4.3)$$

For $g > 1$, $F_g$ takes the following form.

$$F_g = \sum_{\substack{\alpha_j > 1, \\ \beta_i > 1}} \langle \alpha_1 \ldots \alpha_s; \beta_1 \ldots \beta_l | \alpha, \beta, \gamma \rangle_g \frac{M_{\alpha_1} \ldots M_{\alpha_s} J_{\beta_1} \ldots J_{\beta_l}}{M_1^\alpha J_1^\beta d^\gamma}, \qquad g \geq 1, \qquad (4.4)$$

where $d = x - y$, the brackets denote rational numbers and $\alpha$, $\beta$ and $\gamma$ are non-negative integers. The indices $\alpha_1, \ldots, \alpha_s, \beta_1, \ldots, \beta_l$ take values in the interval $[2, 3g - 2]$ and the summation is over sets of indices obeying the following restrictions.

$$s - \alpha \leq 0, \qquad (s - \alpha = 0) \Leftrightarrow (s = \alpha = 0) \qquad (4.5)$$

$$l - \beta \leq 0, \qquad (l - \beta = 0) \Leftrightarrow (l = \beta = 0), \tag{4.6}$$

$$(\alpha - s) + (\beta - l) = 2g - 2, \tag{4.7}$$

$$\sum_{i=1}^{s}(\alpha_i - 1) + \sum_{j=1}^{l}(\beta_j - 1) + \gamma = 4g - 4, \tag{4.8}$$

$$g - 1 \leq \gamma \leq 4g - 4. \tag{4.9}$$

We note that contour integrals like (4.1) and (4.2) are calculated by taking residuals at zero and at infinity. Hence in our formalism we effectively do not need to know the precise location of the support of the eigenvalues, although it is of course essential in order to get a complete understanding of the model. We see that for real $x$ and $y$, $F_g$ will automatically be real while in the case of models with $x$ and $y$ complex, which must of course be defined via analytical continuation, $F_g$ will in general be complex. In the following we investigate the coupling constant space of the one-cut solution to the model (2.1) defined when necessary and possible via analytical continuation. We localize and classify its critical points and show how in the vicinity of any of these one can define a continuum theory using a double scaling prescription. In particular we show that there exist critical points for which $x$ and $y$ are complex but for which we obtain real valued $F_g$'s for all $g$.

## 5 Classification of critical points

This section is devoted to the study of the critical properties of the one cut solution to the model (2.1). Actually the most straightforward way to localize the critical points is by taking a glance at the expressions (4.3) and (4.4) for the higher genera contributions to $F$. It is obvious that there are several circumstances under which the higher genera contributions to the free energy become singular. One possibility is that $M_1$ or $J_1$ or possibly both acquire a zero of some order. This was the only possibility in the case of the usual 1-matrix model and corresponds to the situation where a number of extra zeros accumulates at one or possibly both ends of the support of the eigenvalue distribution. However with the logarithmic term in the action we also have the possibility of $d$ becoming zero. As shown in section 3 the only way in which $d$ can become zero is by $t$ being equal to one. Of course we can also have a situation where both $d = 0$ and and $M_1$ $(=J_1)$ acquire a zero of some order. A third possibility for singular behaviour is that some of the higher moments diverge as we approach a certain point in the coupling constant space. This will be the case if at the critical point $x \to 0$ (or $y \to 0$). For $x \to 0$ we must require that $t \to 0$ in order that $B_1(x, y)$ is well defined. However, the critical point $t = 0$ can be approached in a number of different ways which are all characterized by $x = \mathcal{O}(t)$. That $t = 0$ and $t = 1$ are critical points of the model (2.1) was first noted in reference [4] where the genus zero contribution to the susceptibility was analyzed. Here we will carry out a complete perturbative analysis of the scaling behaviour of the model in the vicinity of its critical points addressing all types of multi-critical behaviour. We will approach the critical points by fixing the coupling constants $\{g_i\}$ at their critical values, $\{g_i^c\}$, and scaling only $t$. As appears from the discussion above the critical points of the model are naturally divided into three types.

**Type 1:** $t \notin \{0, 1\}$, $M_1 = M_2 = \ldots = M_{m-1} = 0$, $M_m \neq 0$, $m \geq 2$,
$J_1 = J_2 = \ldots = J_{n-1} = 0$, $J_n \neq 0$, $1 \leq n \leq m$.

**Type 2:** $t = 1$, $M_1 = M_2 = \ldots = M_{m-1} = 0$, $M_m \neq 0$, $m \geq 0$

**Type 3:** $t = 0$, $M_0 = M_1 = \ldots = M_{m-1} = 0$, $M_m \neq 0$, $m \geq 0$, $J_0 = J_1 = \ldots = J_{n-1} = 0$, $J_n \neq 0$, $1 \leq n \leq m$

Here and in the following it is understood that for moments of the type $M_k(\{g_i\}, t = 0)$ and $J_k(\{g_i\}, t = 0)$, $t$ should be set to zero before the contour integration is carried out. We note that since we have understood that all scaling is associated with $t$ the usual $m$'th multi-critical points of the hermitian 1-matrix model do not immediately fit into this classification. For the usual $m$'th multi-critical points one uses a scaling prescription where $t$ is kept fixed at $t_c = 0$ and the coupling constants $\{g_i\}$ are scaled. As it will appear in the following for $t \notin \{0, 1\}$ both scaling prescriptions lead to the same critical behaviour. For $t_c = 1$, keeping $t$ fixed at $t_c$ does not make sense (cf. to (4.4)). To characterize the different types of critical behaviour let us denote by $\Lambda_R$ the renormalized version of $t$ (i.e. $\Lambda_R \sim (t - t_c)$) and let us introduce the critical index $\gamma_{str}$ by

$$F_g \sim \Lambda_R^{(2-\gamma_{str})(1-g)}, \qquad g > 1. \tag{5.1}$$

## 5.1 Type 1 critical points

Keeping the coupling constants $\{g_i\}$ fixed at their critical values and assuming for a given value of $t$ the branch points of $W(p)$ to be $x$ and $y$ we find expanding the boundary equations $B_1(x, y)$ and $B_2(x, y)$ keeping only leading order terms

$$c_m(x - x_c)^m M_m^c + c_n(y - y_c)^n J_n^c = 0 \tag{5.2}$$
$$c_m(x - x_c)^m x_c M_m^c + c_n(y - y_c)^n y_c J_n^c = (t_c - t) \tag{5.3}$$

where
$$M_m^c = M_m(\{g_i^c\}, t_c), \qquad J_k^c = J_k(\{g_i^c\}, t_c), \qquad c_m = \frac{(2m-1)!!}{m! \, 2^m} \qquad (5.4)$$

If we introduce a renormalized coupling constant $\Lambda_R$ by
$$(t_c - t) = a^m \Lambda_R \qquad (5.5)$$

we see that
$$(x - x_c) \sim a\Lambda_R^{1/m}, \qquad (y - y_c) \sim (a^m \Lambda_R)^{1/n} \qquad (5.6)$$

while of course $d \sim a^0$ since per assumption $t \neq 1$. Furthermore expanding the moments one finds
$$M_k \sim (a\Lambda_R^{1/m})^{m-k}, \quad k \in [1, m], \qquad J_l \sim (a^m \Lambda_R)^{\frac{n-l}{n}}, \quad l \in [1, n]. \qquad (5.7)$$

(We note that in order for (5.7) to be true it is necessary that neither $x$ nor $y$ tend to zero at the critical point. However, as pointed out above $x \to 0$ (or $y \to 0$) is not possible for $t$ finite.) From the expression (3.9) for $\chi_0$ it follows that the leading non analytical term in $F_0$ is given by
$$F_0 = const \cdot \Lambda_R^{2 + \frac{1}{m}} \qquad (5.8)$$

where $const$ is a non universal constant. (We remind the reader that we have assumed that $m \geq n$.) Furthermore from the relation (5.7) it follows that in the scaling limit the genus 1 contribution to the free energy develops a logarithmic singularity of the following type
$$F_1 = -\frac{1}{24}\left\{\frac{m-1}{m} + \frac{n-1}{n}\right\} \log \Lambda_R \qquad 1 < m, \ 1 \leq n \leq m. \qquad (5.9)$$

A given term of $F_g$, $g > 1$ will scale with a negative power of $a$, $P_-^g$, given by
$$P_-^g = \alpha(m-1) + \beta\left(m - \frac{m}{n}\right) - \sum_{i=1}^{s}(m - \alpha_i) - \sum_{j=1}^{l}\left(m - \beta_j \frac{m}{n}\right) \qquad (5.10)$$
$$= (g-1)(2m+2) - \gamma + \left\{(\beta - l) + \sum_{j=1}^{l}(\beta_j - 1)\right\}\left(1 - \frac{m}{n}\right) \qquad (5.11)$$

where to obtain the second equality sign we have made use of the relation (4.8). Now it follows from (4.5), (4.6) and (4.9) that for $n < m$ the dominant terms of $F_g$ are those for which
$$\beta = l = 0, \qquad \gamma = g - 1 \qquad (5.12)$$

These terms hence do not depend on any $J$-moments and have
$$(\alpha - s) = 2g - 2, \qquad \sum_{i=1}^{s}(\alpha_i - 1) = 3g - 3 \qquad (5.13)$$

In deriving (5.11) we implicitly assumed scaling for all $M$ moments involved. By a slight modification of the argument given above it is easy to convince oneself that terms which contain a moment $M_k$ with $k > m$ will be subdominant in the limit $a \to 0$. Hence in this limit $F_g$ can be written as

$$F_g = \sum_{1 < \alpha_j \leq m} \langle \alpha_1 \ldots \alpha_s | \alpha \rangle_g \frac{M_{\alpha_1} \ldots M_{\alpha_s}}{M_1^\alpha d_c^{g-1}} \tag{5.14}$$

All terms in (5.14) scale as given by

$$Max\{P_-^g\} = (g-1)(2m+1) \tag{5.15}$$

and we see the possibility of a double scaling limit emerging. Furthermore, bearing in mind (5.5), we recover the behaviour (5.1) with

$$\gamma_{str} = -1/m. \tag{5.16}$$

For $n = m$ all terms with $\gamma = g - 1$ contribute in the scaling limit and $F_g$ becomes a sum of two terms of the type (5.14), one which involves $M$-moments and one which involves $J$-moments (cf. to reference [7]). Obviously the relations (5.15) and (5.16) hold also in this case. The type 1 multi-critical points of the generalized Penner model clearly belong to the same universality class as the usual $m$'th multi-critical points of the hermitian 1-matrix model.

## 5.2 Type 2 critical points

Let us consider a critical point $(\{g_i\}, t) = (\{g_i^c\}, 1)$ for which $M_1 = M_2 = \ldots = M_{m-1} = 0$, $M_m \neq 0$, $m \geq 1$ and let us approach the point by setting $(\{g_i\}, t) = (\{g_i^c\}, t)$. We note that $M_k^c = J_k^c$ since $x_c = y_c$. Assuming for a given value of $t$ the branch points of $W_0(p)$ to be $x$ and $y$ we find by expanding $B_1(x, y)$ keeping only leading order terms

$$\sum_{i=0}^{m} M_m^c \, c_i \, c_{m-i} (x - x_c)^i (y - y_c)^{m-i} = 0 \tag{5.17}$$

which tells us that $(x - x_c) \sim (y - y_c)$. The expansion of $B_2(x, y)$ leads to the following relation

$$\sum_{i=0}^{m+1} \left\{ x_c \, M_{m+1}^c + M_m \right\} c_i \, c_{m+1-i} \, (x - x_c)^i (y - y_c)^{m+1-i} = t_c - t \tag{5.18}$$

(The vanishing of the terms of order $(x - x_c)^m$ is due to (5.17).) We now define a renormalized coupling constant $\Lambda_R$ by

$$t_c - t = a^{m+1} \Lambda_R \tag{5.19}$$

and get
$$d \sim a\Lambda_R^{1/(m+1)} \tag{5.20}$$

The moments scale in the following way

$$M_k \sim J_k \sim \left(a\Lambda_R^{1/(m+1)}\right)^{m-k}, \qquad k \in [0, m]. \tag{5.21}$$

Using the expression (3.9) it is easy to see that the singular behaviour at genus zero is caused by the scaling of $d$ and that

$$F_0 = -\frac{1}{m+1}\Lambda_R^2 \log \Lambda_R \tag{5.22}$$

as was also found in reference [4]. Furthermore using (4.3), (5.20) and (5.21) we see that in the scaling limit we have *independently of m*

$$F_1 = -\frac{1}{12}\log \Lambda_R. \tag{5.23}$$

This property of the multi-critical Penner models was conjectured (but not proven) in reference [6]. Here we can address also the question of the behaviour of these models for $g > 1$. A given term of $F_g$, $g > 1$ will scale with a negative power of $a$, $P_-^g$ given by

$$\begin{aligned}
P_-^g &= ((\alpha - s) + (\beta - l))(m - 1) + \sum_{i=1}^{s}(\alpha - 1) + \sum_{j=1}^{s}(\beta - 1) + \gamma \tag{5.24}\\
&= (g - 1)(2m + 2) \tag{5.25}
\end{aligned}$$

Here we find the amazing result that all terms in our expression (4.4) are potentially relevant for the scaling limit. By potentially relevant we mean relevant for $m$ sufficiently large: We have in deriving (5.24) and (5.25) implicitly assumed that all moments scale. A closer analysis reveals that as in the previous case terms which contain a moment $M_k$, $k > m$ can be neglected in the limit $a \to 0$. Now bearing in mind the relation (5.19) we see that independently of $m$ we have

$$\gamma_{str} = 0. \tag{5.26}$$

Among the type 2 critical points we find the critical point of the Penner model, $g_1^c = -1$, $g_i^c = 0$, $i > 1$. [1] It is well known [2] that this model exhibits the scaling behaviour characteristic of a theory describing 2D quantum gravity interacting with matter with central charge, $c = 1$. However, it appears that this scaling behaviour occurs for all type 2 critical points with $M_1^c \neq 0$. Let us note that for these models all scaling of the higher genera contributions to the free energy is due to the scaling of $d$ and the terms which are important in the limit $a \to 0$ are those for which

$$\gamma = 4(g-1), \qquad s = l = 0, \qquad \alpha + \beta = 2(g-1) \tag{5.27}$$

---

[1] For the Penner model one actually has to keep next to leading order terms in the expansion of the boundary equations. However, it its easy to show that the conclusions stated above hold also in this case.

In particular these terms do not depend on any moments other than $M_1$ and $J_1$.

## 5.3 Type 3 critical points

For simplicity let us consider a point for which $J_1 \neq 0$. The generalization to the case $J_1 = J_2 = \ldots = J_{n-1} = 0$, $J_n \neq 0$, $1 \leq n \leq m$ is straightforward although more tedious than in the case of the type 1 multi critical models. (We remind the reader that for moments of the type $M_k(\{g_i\}, t=0)$, $J_k(\{g_i\}, t=0)$ it is understood that $t$ should be set to zero before the contour integration is carried out.) Again we keep the coupling constants $\{g_i\}$ fixed at their critical values and assume at a given value of $t$ the branch points of $W(p)$ to be $x$ and $y$. After performing a few rearrangements we can write the boundary equations (3.2) and (3.3) as

$$c_m x^m (x\, y_c)^{1/2} M_m^c + c_1 (y - y_c) J_1^c y_c = 0 \tag{5.28}$$

$$c_m x^m (x\, y_c)^{1/2} M_m^c + t = 0 \tag{5.29}$$

Hence if we introduce a renormalized coupling constant $\Lambda_R$ by

$$t = a^{m+1/2} \Lambda_R \tag{5.30}$$

we have

$$x \sim a \Lambda_R^{\frac{1}{m+1/2}}, \qquad (y - y_c) \sim a^{m+1/2} \Lambda_R \tag{5.31}$$

Making use of the expression (3.9) for $\chi_0$ we find

$$F_0 = \frac{1}{2m+1} \Lambda_R^2 \log \Lambda_R. \tag{5.32}$$

For the moments we have

$$M_k \sim a^{m-k}, \qquad k \in [0, \infty] \tag{5.33}$$

while the $J$-moments do not scale. Hence from (4.3) we see that the genus 1 contribution to the free energy behaves in the scaling limit as

$$F_1 = -\frac{1}{24} \left( \frac{m-1}{m+1/2} \right) \log \Lambda_R. \tag{5.34}$$

To handle the higher genera contributions to $F$ we note that except for the fact that in the present case all $M$-moments scale the behaviour (5.33) is the same as that of a type 1 multi-critical model with $(m,n) = (m,1)$. Hence we have the same value of $max\{P_-^g\}$ as in that case which, bearing in mind the relation (5.31) tells us that

$$\gamma_{str} = 0. \tag{5.35}$$

## 6 The fermionic 1-matrix model

In this section we consider the following 1-matrix model

$$Z = e^{N^2 F} = \int d\bar{\Psi} d\Psi \exp\left(-N \operatorname{Tr} V(\bar{\Psi}\Psi)\right) \tag{6.1}$$

where $\bar{\Psi}$ and $\Psi$ are $N \times N$ matrices whose matrix elements are independent Grassmann variables. We take the interaction to be of the following type

$$V(\bar{\Psi}\Psi) = \sum_{k=1}^{\infty} \frac{g_k}{k} (\bar{\Psi}\Psi)^k. \tag{6.2}$$

For this model we define the 1-loop correlator by

$$W(p) = \langle \frac{1}{N} \operatorname{Tr} \frac{1}{p - \bar{\Psi}\Psi} \rangle \tag{6.3}$$

and the genus expansion of the free energy and the 1-loop correlator looks as in equation (2.4). To derive the loop equations of the model it is convenient to consider a transformation of the fields $(\bar{\Psi}, \Psi)$ of the following type

$$\Psi \rightarrow \Psi + \epsilon \frac{\Psi}{p(p - \bar{\Psi}\Psi)}, \tag{6.4}$$

$$\bar{\Psi} \rightarrow \bar{\Psi}. \tag{6.5}$$

Under such a transformation the measure changes as [2]

$$d\bar{\Psi} d\Psi \rightarrow d\bar{\Psi} d\Psi \left(1 - \epsilon \left[\left(\operatorname{Tr}\left(\frac{1}{p - \bar{\Psi}\Psi}\right)\right)^2 - 2\operatorname{Tr}\left(\frac{1}{p(p - \bar{\Psi}\Psi)}\right)\right]\right). \tag{6.7}$$

---

[2] In case we were considering in stead of the matrices $(\bar{\Psi}, \Psi)$ with Grassmannian matrix elements complex matrices $(\phi^\dagger, \phi)$ a similar shift of variables would lead to the following change of the measure [11]

$$d\phi^\dagger d\phi \rightarrow d\phi^\dagger d\phi \left(1 + \epsilon \left(\operatorname{Tr}\left(\frac{1}{p - \phi^\dagger \phi}\right)\right)^2\right). \tag{6.6}$$

What causes the difference between this formula and formula (6.7) is the fact that in comparison with the complex matrix model the fermionic matrix model has an additional factor $(-1)$ appearing in the change of the measure for transformations of the type $\Psi \rightarrow \Psi + \epsilon \Psi (\bar{\Psi}\Psi)^n$, $n \geq 1$.

Using the result (6.7) it is easy to show that the invariance of the partition function under the transformation (6.4), (6.5) leads to the following loop equation

$$\frac{1}{p}\oint_C \frac{d\omega}{2\pi i} \frac{\omega V'(\omega)}{p-\omega} W(\omega) + \frac{2}{p} W(p) = (W(p))^2 + \frac{1}{N^2} \frac{d}{dV(p)} W(p) \tag{6.8}$$

where the notation is as in section 2. This equation is identical to equation (2.6) for $t = 2$ [8]. Hence we learn that when analytically continued to $t = 2$ the hermitian matrix model (2.1) is equivalent to the fermionic matrix model (6.1). In particular we know from section 3 and section 4 the complete perturbative one-cut solution to the model (6.1).

The boundary equations for $t = 2$ reveal an interesting feature of the fermionic model. Inserting $t = 2$ in (3.2) and (3.3) we find

$$B_1(x,y) = \oint \frac{d\omega}{2\pi i} \frac{V'(\omega)}{\sqrt{(\omega-x)(\omega-y)}} - \frac{2}{(xy)^{1/2}} = 0, \tag{6.9}$$

$$B_2(x,y) = \oint \frac{d\omega}{2\pi i} \frac{\omega V'(\omega)}{\sqrt{(\omega-x)(\omega-y)}} = 0. \tag{6.10}$$

We see that the roles of the boundary equations are in a sense opposite to what was the case for the ordinary hermitian 1-matrix model. The concepts symmetrical and non symmetrical are interchanged, i.e. an odd potential leads to the branch points of $W(p)$ being placed symmetrically with respect to the origin while an even potential leads to a non symmetrical placing of the branch points. Let us consider the case of an odd potential, i.e.

$$V(\bar{\Psi}\Psi) = \sum_{k=0}^{\infty} \frac{g_{2k+1}}{2k+1} (\bar{\Psi}\Psi)^{2k+1}. \tag{6.11}$$

In this case, if we set $x = -y = z$ the boundary equation (6.10) is trivially fulfilled, since there is no residue at infinity. The remaining boundary equation reads

$$\sum_{j=0}^{\infty} g_{2j+1}\, c_j\, z^{2j} - \frac{2}{(-z^2)^{1/2}} = 0 \tag{6.12}$$

which admits a solution of the type $z = i\mu$. In order to recover in the limit $g_{2j+1} \to 0$, $j > 0$ the previously obtained results for the Penner model we should take the negative square root of $(-z^2)^{1/2}$. However, if we want to recover in the limit $g_{2j+1} \to 0$, $j > 0$ a gaussian model we should take the positive square of $(-z^2)^{1/2}$. In the following we will consider the latter possibility since we wish to have an interpretation of our model in terms of random surfaces. In any case we see that for a fermionic matrix model with odd potential we have for the branch points of $W(p)$, $x = -y = i\mu$. This has the following implications for the moments

$$M_n = (-1)^n J_n. \tag{6.13}$$

It is obvious that the fermionic model possesses only critical points of type 1. To investigate the scaling behaviour in the vicinity of these points we will set $\{g_i\} = \{g \cdot g_i^c\}$ and scale $g$ to 1. As mentioned earlier for type 1 critical points the critical behaviour is the same whether the scaling is associated with $t$ or $g$. Hence the critical points of the fermionic model belong to the same universality class as the critical points of the usual hermitian 1-matrix model. The value of $\gamma_{str}$ for the $m$'th multi-critical fermionic model coincides with the value of $\gamma_{str}$ for the $m$'th multi-critical hermitian model to all genera. The critical indices being identical does not mean that the topological expansion in the vicinity of a $m$'th multi-critical point is the same for the fermionic and the hermitian 1-matrix model, however. For the fermionic 1-matrix model the branch points of $W(p)$, i.e. $x$ and $y$ will in general be complex and hence $F_g$ will in general be a complex quantity. In the case where the potential of the fermionic model has a definite parity it is easy to investigate how the branch points being complex affects the topological expansion.

Let us to begin with consider the case of an odd potential. Inserting $x = -y = i\mu$ in the definitions (4.1) and (4.2) we find

$$M_n = \frac{1}{i^n} \tilde{M}_n, \qquad J_n = \frac{1}{i^n} \tilde{J}_n \tag{6.14}$$

where $\tilde{M}$ is real and given by

$$\tilde{M}_n = \oint \frac{d\omega}{2\pi i} \frac{\sum_{j=0}^{\infty} g_{2j+1}\, \mu^{2j}(-1)^j}{(\omega-\mu)^{n+1/2}(\omega+\mu)^{1/2}} - \frac{2}{(-\mu)^n \mu} \tag{6.15}$$

and $\tilde{J}_n = (-1)^n \tilde{M}_n$ (cf. to equation (6.13)). In addition we of course have

$$d = 2i\mu = i\tilde{d} \tag{6.16}$$

where $\tilde{d}$ is real. Inserting the relations (6.14) and (6.16) into the expression (4.4) we find for a given term $f_g$ of our expression (4.4) for the genus $g$ contribution to the free energy

$$f_g = (i)^{\alpha+\beta-\sum_{i=1}^{s}\alpha_i-\sum_{j=1}^{l}\beta_j-\gamma}\, \tilde{f}_g \tag{6.17}$$

where $\tilde{f}_g$ is real and appears from $f_g$ in (4.4) by replacing $M$, $J$ and $d$ by $\tilde{M}$, $\tilde{J}$ and $\tilde{d}$. Using the relations (4.7) and (4.8) one finds that

$$F_g = (-1)^{g-1} \tilde{F}_g. \tag{6.18}$$

Now it seems as if a possibility of a topological expansion with alternating signs emerges. To test the viability of this possibility we must of course study in detail the $\tilde{M}$'s. Let us do so restricting ourselves to the vicinity of a $m = 2$ multi-critical point.

At this point it is instructive to recall what the situation was like for the hermitian 1-matrix model without the logarithmic term. To be specific let us consider the model where only $g_1$ and $g_3$ are different from zero. For this model only $M_1$ and $M_2$ (and $J_1$ and $J_2$) are non vanishing (where we now understand that the moments are defined without the $t$ term). The support of the eigenvalue distribution (in the perturbative region of the coupling constant space) is asymmetric and we denote it as $[y, x]$. The model has a $m = 2$ multi-critical point, namely

$$g_1^c = \frac{3}{2^{2/3}}, \qquad g_3^c = -1. \tag{6.19}$$

Let us approach this critical point by setting $g_i = g \cdot g_i^c$. Expanding the boundary equations in powers of $(x - x_c)$ and $(y - y_c)$ we recover the equations (5.2) and (5.3) with $(m, n) = (2, 1)$. In particular inserting (5.2) in (5.3) we find

$$(1 - g) = c_2 (x - x_c)^2 M_2^c. \tag{6.20}$$

In the scaling limit all dependence on $J$-moments disappear and for the $M$-moments we have

$$M_2 = M_2^c = g_3^c < 0, \tag{6.21}$$
$$M_1 = \frac{3}{2} M_2^c (x - x_c) > 0. \tag{6.22}$$

Furthermore for $d$ we have

$$d = x_c - y_c = d_c > 0. \tag{6.23}$$

Now let us return to the fermionic model with odd potential and let us try to localize a $m = 2$ critical point, i.e. a point for which $M_0 = M_1 = 0$. If we assume for simplicity that only $g_1$ and $g_3$ are different from zero we get the following equations for $g_1^c$, $g_3^c$ and $\mu_c$

$$M_0^c = -\frac{1}{2} g_3^c \mu_c^3 + g_1^c \mu_c - 2 = 0, \tag{6.24}$$
$$M_1^c = g_3^c \mu_c^3 - 2 = 0. \tag{6.25}$$

It is easy to see that the following solution is possible

$$g_3^c = 1, \qquad g_1^c = \frac{3}{2^{1/3}}, \qquad \mu_c = 2^{1/3}. \tag{6.26}$$

As before we approach the critical point by setting $g_i = g \cdot g_i^c$ and expand our (single) boundary equation in powers of $z - z_c$ (where $z = i\mu$). After a few rewritings where only leading order terms are kept we can represent our boundary equation in the following form

$$(g - 1) = c_2 \, M_2^c \, \mu_c \, (\mu - \mu_c)^2 \tag{6.27}$$

which looks completely similar to the corresponding equation of the ordinary hermitian 1-matrix model (cf. to (6.20)). In this case we find in the scaling limit for the moments of interest

$$\tilde{M}_2 = -(g_3^c + \frac{2}{\mu_c^3}) = -2g_3^c < 0, \tag{6.28}$$

$$\tilde{M}_1 = \tilde{M}_2^c(\mu - \mu_c) = -2g_3^c(\mu - \mu_c) > 0. \tag{6.29}$$

Furthermore we of course have

$$\tilde{d} = 2\mu_c > 0. \tag{6.30}$$

Since $\tilde{M}_1$, $\tilde{M}_2$ and $\tilde{d}$ have the same signs as $M_1$, $M_2$ and $d$ for the hermitian model we get for the fermionic $m = 2$ multi-critical model a topological expansion with alternating signs. Apart from the signs the topological expansion of the fermionic model is identical to the usual Painlevé expansion (for suitable normalization of the cosmological constant). The question of the possible change of signs in the topological expansion of the higher multi-critical fermionic models requires further investigation. However, on the basis of the above presented example we put forward the conjecture that for all $m$'th multi-critical points of the fermionic 1-matrix model with odd potential we have in the scaling limit for the genus $g$ contribution to the free energy, $F_g^{ferm}$

$$F_g^{ferm} = (-1)^{g-1} \, F_g^{herm} \tag{6.31}$$

where $F_g^{herm}$ is the genus $g$ contribution to the free energy of a $m$'th multi-critical hermitian model obtained from a symmetrical potential.

Let us close this section by commenting on the fermionic matrix model with an even potential. In this case an analysis of the boundary equations shows that the branch points of $W(p)$ are again imaginary. However, now they both lie on either the positive or the negative real axis. (If the potential does not have a definite parity the branch points will in general have a real as well as an imaginary part.) Also for the symmetric fermionic matrix model we get a set of relations like (6.14). However, the prefactor $\frac{1}{i^n}$ is replaced by $\frac{1}{i^{n+1}}$. This modification results in (6.18) being replaced by

$$F_g = \tilde{F}_g \tag{6.32}$$

Analyzing in detail the $m = 2$ multi-critical point is in this case a lot more complicated than in the case of an odd potential, because the cut is no longer symmetric and the moments involve more terms due to the larger degree of the potential. Of course the analysis of the higher $m$'th multi-critical points is even more difficult. However, we put forward the conjecture that the topological expansion in the vicinity of a $m$'th multi-critical point obtained from a fermionic matrix model with even potential coincides with the topological expansion in the vicinity of a $m$'th multi-critical point obtained from a non symmetrical hermitian matrix model.

# 7  Conclusion

The combination of the loop equations and the moment description constitute a powerful tool for studying higher genera contributions in matrix models. In particular the moment variables make it very simple to localize critical points and to investigate the scaling behaviour in the vicinity of these points.

Previously the moment technique was developed only for 1-matrix models with polynomial potentials. Here we have extended the method to the case where in addition a non polynomial term, namely a logarithm, appears in the interaction, i.e. to the generalized Penner model. The investigation of the coupling constant space of the generalized Penner model revealed no unknown types of scaling behaviour. Our solution of the generalized Penner model provided us with a solution of the fermionic 1-matrix model and the possible types of critical behaviour of this model turned out to be characterized by the same value of $\gamma_{str}$ as the $m$'th multi-critical points of the usual hermitian 1-matrix model. A feature of the fermionic 1-matrix model which deserves some attention, though, is that it has a $m = 2$ multi-critical point for which the topological expansion has alternating signs but otherwise coincides with the usual Painlevé expansion. Having alternating signs this series might very well be Borel summable. It would be interesting if the fermionic nature of the model could be given a world sheet interpretation.

One interesting prospect of the present work is the possibility of generalizing the moment technique to the hermitian two-matrix model which is known to be capable of describing all minimal conformal models coupled to gravity [12]. The study of loop equations for the hermitian two-matrix model has been initiated [13], but explicit results for correlators are few and limited to genus zero. Another promising application of our results is the genus zero analysis of matrix models in dimensions, $D > 1$ [14]. Here the task would be to localize critical points and devise possible continuum limits. The genus zero loop equations were obtained in reference [15]. Of particular interest is the hermitian matrix model with a logarithmic potential for which these equations can be exactly solved for any $D$ [16].

## Appendix A  The eigenvalue support for various $t$

Let us investigate in some detail the question of the location of the support of the eigenvalue distribution for the Penner model

$$U(\phi) = -\phi + t\log\phi. \tag{A.1}$$

From equation (3.14) and (3.15) we know explicitly the position of the endpoints $x$ and $y$ of the distribution as a function of $t$. As explained earlier the exact location of the support can be found only by studying the function $G(\lambda)$ defined by (3.5). The support is an arc connecting $y$ to $x$ along which $G(\lambda)$ is purely imaginary and which is embedded in a region where $\text{Re}(G(\lambda)) < 0$. Using our solution (3.1) we can write down an explicit expression for $G(\lambda)$. We find

$$\begin{aligned} G(\lambda) &= -\int_y^\lambda \frac{1}{p}(p-x)^{1/2}(p-y)^{1/2} \\ &= -\Big[(p-x)^{1/2}(p-y)^{1/2} + (2-t)\log\Big(2\big[(p-x)^{1/2}(p-y)^{1/2} + p + (2-t)\big]\Big) \\ &\quad + t\log\left(\frac{2\big[-t\,(p-x)^{1/2}(p-y)^{1/2} + p(2-t) + t^2\big]}{p}\right)\Big]_y^\lambda \end{aligned} \tag{A.2}$$

where we have on the way made use of the boundary equations (3.11) and (3.12).

Below we have shown the sign variation of $\text{Re}(G(\lambda))$ in some spectacular cases assuming the cut of the square root $(p-x)^{1/2}(p-y)^{1/2}$ to be the straight line connecting $x$ and $y$. Before proceeding to analyzing these plots let us note certain characteristica of the function $\text{Re}(G(\lambda))$ which can easily be read off from (A.2). First we see that as $p \to \pm\infty$ we have $\text{Re}(G(\lambda)) \to \mp\infty$ and as $p \to 0$ we have $\text{Re}(G(\lambda)) \to +\infty$ (provided $t \neq 0$). Futhermore it is easy to convince oneself that if $x$ and $y$ are real $\text{Re}(G(\lambda)) = 0$ for $\lambda \in [y,x]$. However, if $x$ and $y$ are imaginary we have $\text{Re}(G(\lambda)) \neq 0$ for $\lambda$'s on the imaginary axis between $x$ and $y$.

Figure 6 shows the case $t = 1$. We see that the branch cut $[y,x]$ has collapsed to a point and that there is a loop encircling the point $\lambda = 0$ along which $\text{Re}(G(\lambda)) = 0$. Inside the loop $\text{Re}(G(\lambda)) > 0$. The appearance of such a loop going through the point $x$ is a common feature for all $t < 1$, $t \neq 0$. For $t < 0$ the support of the eigenvalue distribution is simply $[y,x]$. However, for $0 < t < 1$ with the cut of $W(p)$ only reaching from $y$ to $x$ we have actually not fulfilled the requirement that we should take the negative square root of $(xy)^{1/2}$ (cf. to section 3). However, this requirement will be fulfilled if we deform our original cut so that it includes also the loop. Then we will have $\text{Re}(G(\lambda)) < 0$ both inside and outside the loop and hence the loop belonging to the support of the eigenvalue distribution. When $t > 1$ the points $x$ and $y$ become complex and each others complex conjugate. In figure 7 we have plotted $\text{Re}(G(\lambda))$ for $t = \frac{3}{2}$. We see that there is still an arc connecting $x$ and $y$ along which $\text{Re}(G(\lambda)) = 0$. This arc passes the real axis to the right of the origin. (We note that the discontinuity of $\text{Re}(G(\lambda))$ along the vertical line connecting $x$ and $y$ of course has its origin in the cut of the square root $(p-x)^{1/2}(p-y)^{1/2}$ being placed along this line). As in the case of $0 < t < 1$ we can now deform our original cut into the arc with $\text{Re}(G(\lambda)) = 0$ connecting $x$ and $y$, thereby identifying this arc with the support of the eigenvalue distribution and fulfilling the requirement concerning the sign of $(xy)^{1/2}$. Finally in figure 8 we have plotted $\text{Re}(G(\lambda))$ in the fermionic case $t = 2$. The figure examplifies the statement that the roles of even and odd potentials are interchanged for the fermionic 1-matrix model. We see for the odd potential $V(\phi) = -\phi$ a completely symmetric picture. The analysis of the cut structure is left to the reader.

# References

[1] V.A. Kazakov, Mod. Phys. Lett. **A4** (1989) 2125.

[2] J. Distler and C. Vafa, Mod. Phys. Lett. A6 (1991) 259

[3] R.C. Penner, J. Diff. Geom. 27 (1988) 35.

[4] C. Tan, Mod. Phys. Lett. A6 (1991) 1373

[5] S. Chaudhuri, H. Dykstra and J. Lykken, Mod. Phys. Lett A6 (1991) 1665

[6] C. Tan, Phys. Rev. D8 (1992) 2862

[7] J. Ambjørn, L. Chekhov, C.F. Kristjansen and Yu. Makeenko, Nucl. Phys. B404 (1993) 127

[8] Yu. Makeenko and K. Zarembo, *Adjoint Fermion Matrix Models*, ITEP-YM-5-93

[9] E. Brézin, C. Itzykson, G. Parisi and J.B. Zuber, Commun. Math. Phys. 59 (1978) 35.

[10] F. David, Nucl. Phys. B348 (1991) 507

[11] J. Ambjørn, J. Jurkiewicz and Yu. Makeenko, Phys. Lett. B251 (1990) 517.

[12] M. Douglas, in *Random surfaces and quantum gravity (Cargèse 1990)*, eds. O. Alvarez et. al., Plenum, 1991, p.77;
J.-M. Daul, V.A. Kazakov and I.K. Kostov, Nucl. Phys. B409 (1993) 311

[13] J. Alfaro and J.C. Retamal, Phys. Lett. B222 (1989) 429;
E. Gava and K.S. Narain, Phys. Lett. B263 (1991) 213;
J. Alfaro, Phys. Rev. D47 (1993) 4714;
M. Staudacher, Phys. Lett. B305 (1993) 332

[14] V.A. Kazakov and A.A. Migdal, Nucl. Phys. B397 (1993) 214

[15] Yu. Makeenko, Mod. Phys. Lett. A8 (1993) 209;
     M.I. Dobroliubov, Yu. Makeenko and G.W. Semenoff, Mod. Phys. Lett. A8 (1993) 2387
[16] Yu. Makeenko, Phys. Lett. B314 (1993) 197

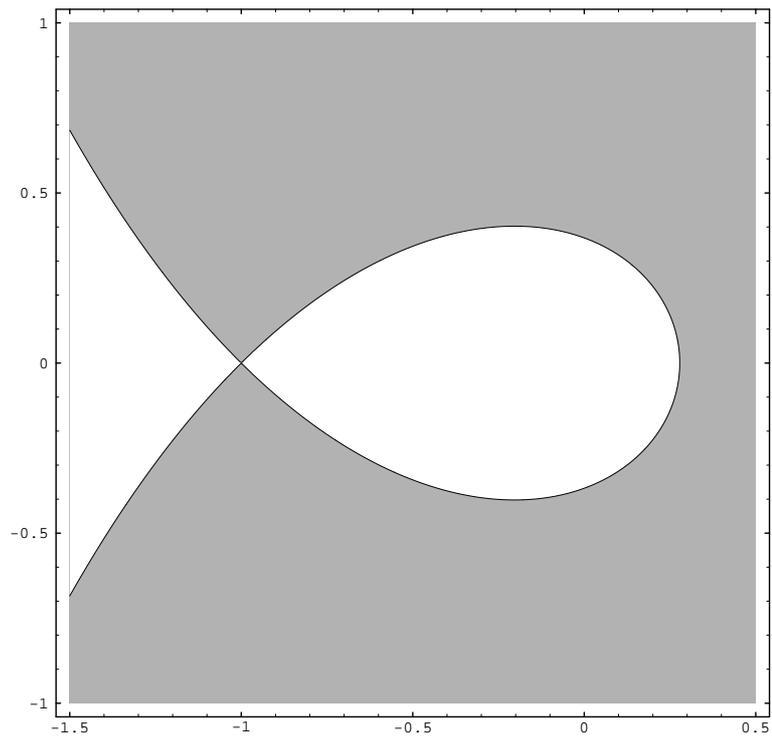

Figure 6: The sign variation of $\mathrm{Re}(G(\lambda))$ for $t = 1$. $\mathrm{Re}(G(\lambda)) < 0$ in the gray region and $\mathrm{Re}(G(\lambda)) > 0$ in the white region.

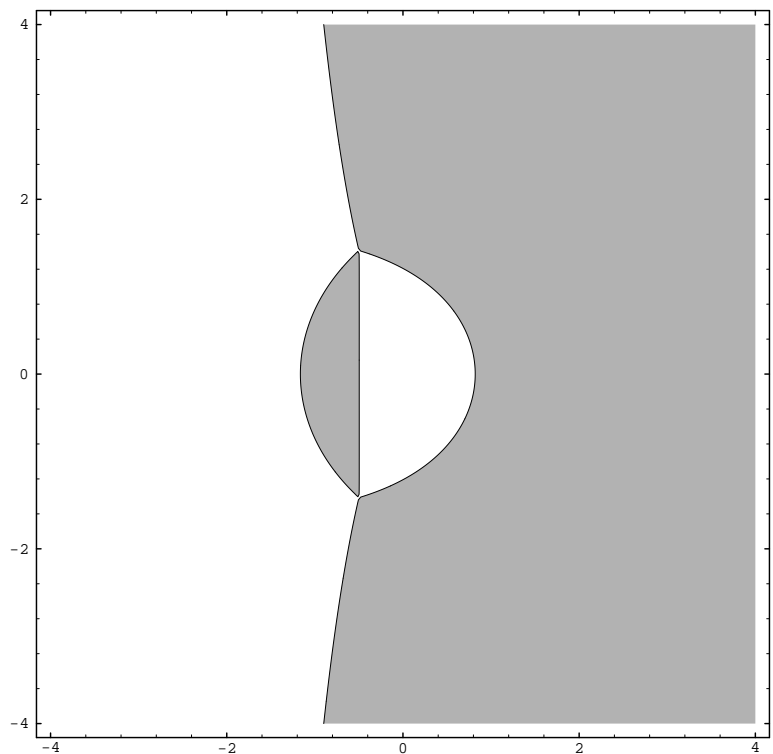

Figure 7: The sign variation of $\text{Re}(G(\lambda))$ for $t = 3/2$. $\text{Re}(G(\lambda)) < 0$ in the gray region and $\text{Re}(G(\lambda)) > 0$ in the white region.

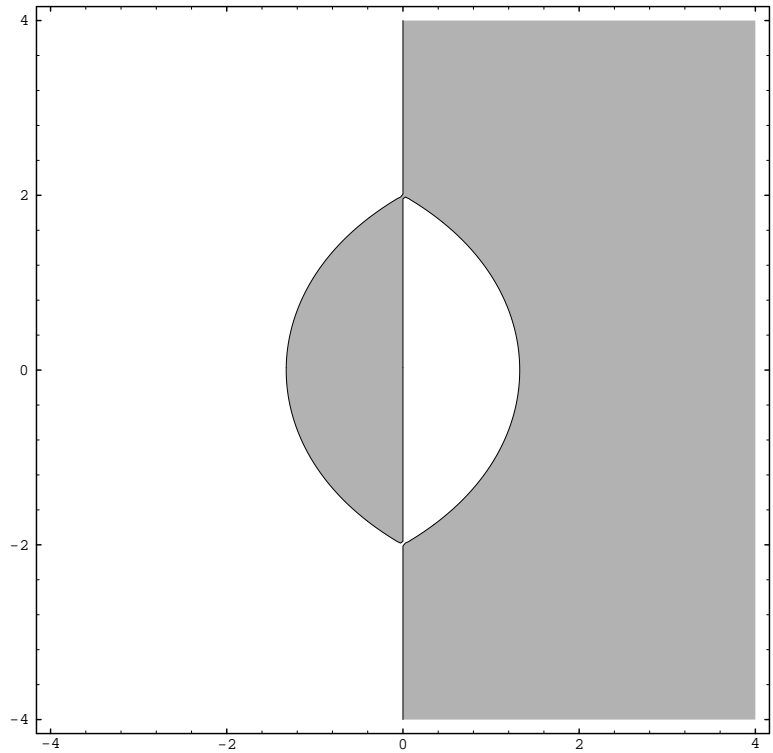

Figure 8: The sign variation of $\text{Re}(G(\lambda))$ for $t = 2$. $\text{Re}(G(\lambda)) < 0$ in the gray region and $\text{Re}(G(\lambda)) > 0$ in the white region.